# Noise Reduction in Diffusion MRI Using Non-Local Self-Similar Information in Joint $x$-$q$ Space

Geng Chen[a], Yafeng Wu[c], Dinggang Shen[a,b,*], Pew-Thian Yap[a,*]

[a] *Department of Radiology and Biomedical Research Imaging Center (BRIC) University of North Carolina at Chapel Hill, NC, U.S.A.*
[b] *Department of Brain and Cognitive Engineering, Korea University, Seoul, Korea*
[c] *Data Processing Center, Northwestern Polytechnical University, Xi'an, China*

## Abstract

Diffusion MRI affords valuable insights into white matter microstructures, but suffers from low signal-to-noise ratio (SNR), especially at high diffusion weighting (i.e., $b$-value). To avoid time-intensive repeated acquisition, post-processing algorithms are often used to reduce noise. Among existing methods, non-local means (NLM) has been shown to be particularly effective. However, most NLM algorithms for diffusion MRI focus on patch matching in the spatial domain (i.e., $x$-space) and disregard the fact that the data live in a combined 6D space covering both spatial domain and diffusion wavevector domain (i.e., $q$-space). This drawback leads to inaccurate patch matching in curved white matter structures and hence the inability to effectively use recurrent information for noise reduction. The goal of this paper is to overcome this limitation by extending NLM to the joint $x$-$q$ space. Specifically, we define for each point in the $x$-$q$ space a spherical patch from which we ex-tract rotation-invariant features for patch matching. The ability to perform patch matching across $q$-samples allows patches from differentially orientated structures to be used for effective noise removal. Extensive experiments on synthetic, repeated acquisition, and real data demonstrate that our method outperforms state-of-the-art methods, both qualitatively and quantitatively.

*Corresponding authors: ptyap@med.unc.edu (Pew-Thian Yap), dgshen@med.unc.edu (Dinggang Shen)

## 1. Introduction

Diffusion MRI (DMRI) relies on its sensitivity to the displacement of water molecules to probe tissue microstructure. To be able to characterize fine microstructural details, the diffusion weighting (i.e., $b$-value) needs to be sufficiently high, allowing for example more accurate separation of fiber bundles crossing at small angles and greater sensitivity to the restricted diffusion of water molecules trapped inside axons. However, due to the significant attenuation of the MR signal at high diffusion weightings, the low signal-to-noise ratio (SNR) poses significant challenges to subsequent analysis.

A straightforward means to improve SNR is by repeating and averaging scans Johansen-Berg and Behrens (2013), which however inevitably prolongs acquisition times and is hence impractical in clinical settings. In view of this, post-acquisition denoising methods have been widely adopted Wiest-Daesslé et al. (2007, 2008); Descoteaux et al. (2008); Becker et al. (2012); Manjón et al. (2013); Becker et al. (2014); Lam et al. (2014); Yap et al. (2014); Varadarajan and Haldar (2015); Veraart et al. (2016); St-Jean et al. (2016). Among existing methods, non-local means (NLM) has been shown to be particularly good at preserving edges when reducing noise. NLM avoids blurring by averaging over recurrent image patterns collected via patch matching.

NLM has been applied to reducing noise in DMRI data Wiest-Daesslé et al. (2007, 2008); Descoteaux et al. (2008); Yap et al. (2014). Existing NLM methods denoise DW images as individual images, a multi-spectral vector image, or parametric maps given by a diffusion model Wiest-Daesslé et al. (2007). However, these methods mainly focus on patch matching in the spatial domain (i.e., $x$-space), despite the fact that DMRI data live in a combined space consisting of both spatial $x$-space and diffusion wavevector $q$-space. This causes NLM to be less effective in locating self-similar patterns in highly curved white matter structures, resulting in smoothing artifacts caused by averaging over differentially oriented structures. Another limitation of NLM is the *rare patch effect* Duval et al. (2011); Deledalle et al. (2012); Salmon and Strozecki (2012). This phenomenon happens when matching structures cannot be found, leading to the degradation of fine details and causing halos around object boundaries. A natural solution to this problem is to expand the search extent Prima and Commowick (2013); Chen et al. (2016b) so that the possibility of finding matching structures can be increased. However, this significantly increases computation time and might result in false-positive matches.

To overcome these limitations, in this paper we extend NLM beyond $x$-space to include $q$-space for improved denoising in DMRI. Specifically, for each point in $x$-$q$ space, we first define a patch covering a $q$-space neighborhood. We then perform patch matching in $x$-$q$ space and assign a weight, indicating neighborhood similarity, for each pair of points in the joint space. Finally, the denoised signal at each point in $x$-$q$ space is estimated via weighted averaging.

The advantage afforded by this extension is fourfold: (i) Non-local information can now be harnessed not only over $x$-space, but also over $q$-space, allowing information to be borrowed across diffusion-weighted (DW) images for effective denoising; (ii) Information from structures oriented in different directions can be used more effectively for denoising without introducing artifacts; (iii) Patch matching complexity can be significantly reduced by leveraging the fact that diffusion signal profiles generally have smooth and simpler shapes; (iv) The simpler shapes also imply that better patch matches can be found more easily, therefore mitigating the rare patch effect.

Comprehensive experiments on synthetic, repeated acquisition, and real data demonstrate that $x$-$q$ space non-local means (XQ-NLM) removes noise effectively while preserving structures and improves the quality of derived quantities, such as fractional anisotropy (FA), orientation distribution function (ODF), and fiber tracts. We compared our method with state-of-the-art methods, including adaptive NLM (ANLM) Manjón et al. (2010), non-local spatial and angular matching (NLSAM) St-Jean et al. (2016), and Marchenko-Pastur principle component analysis (MPPCA) Veraart et al. (2016). Experimental results confirm that our method consistently gives the best performance both qualitatively and quantitatively.

A preliminary version of this work has been presented at a conference Chen et al. (2016a). In this journal version, we (i) extend our method to work with a wider range of noise types resulting from multi-coil MRI and different methods of magnitude signal reconstruction, (ii) perform quantitative evaluation using a ground truth generated using repeatedly acquired data,

(iii) include new results for synthetic and real data as well as additional metrics, covering voxel- and tract-based assessments, (iv) compare our method with state-of-the-art denoising methods (i.e., NLSAM and MPPCA), and (v) include additional discussions that are not part of the conference publication.

The rest of the paper is organized as follows. We will first flesh out in Section 2 the key components of the proposed method. We will then demonstrate the effectiveness of our method in the Section 3 using synthetic and real data. Additional discussions are provided in Section 4 before concluding the paper in Section 5.

## 2. Methods

### *2.1. Noise Adaptation*

#### *2.1.1. Noise Types in Multi-Coil MRI*

The noise distribution of the composite magnitude signal (CMS) Aja-Fernández and Vegas-Sánchez-Ferrero (2016) given by modern multi-coil MRI techniques is dependent on how the $k$-space signal is sampled and how it is used to reconstruct the magnitude signal. For practical purposes, it is commonly assumed that the $k$-space noise is a zero-mean stationary Gaussian process with equal variance in both real and imaginary parts.

Sum of squares (SoS) and spatial matched filter (SMF) are widely used for CMS reconstruction. For $N$ coils with uncorrelated noise, SoS reconstruction from fully-sampled $k$-space data leads to spatially stationary nc-$\chi$ noise distribution with $2N$ degrees of freedom, whereas SMF leads to spatially stationary Rician noise distribution Dietrich et al. (2008). If noise is correlated across coils, the noise distribution of the reconstructed CMS becomes spatially non-stationary.

To accelerate acquisition, parallel MRI subsamples the $k$-space. Two widely used methods are sensitivity encoding (SENSE) Pruessmann et al. (1999) and generalized autocalibrating partially parallel acquisition (GRAPPA) Griswold et al. (2002). The CMS reconstructed by SENSE follows a non-stationary Rician distribution Aja-Fernández et al. (2014). The noise of the CMS reconstructed from GRAPPA data using SoS and SMF follows the non-stationary nc-$\chi$ and Rician distributions, respectively Aja-Fernández et al. (2014).

### *2.1.2. Signal Transformation*

Before denoising, we transform the CMS so that its noise becomes Gaussian distributed, similar to St-Jean et al. (2016). This involves estimating the location parameter and Gaussian noise standard deviation of the nc-$\chi$ distribution and then performing signal transformation using the nc-$\chi$ cumulative distribution function (CDF) and the inverse Gaussian CDF Koay et al. (2009a). Such signal transformation reduces the complexity of the denoising algorithm by not having to deal with the nc-$\chi$ nature of the noise St-Jean et al. (2016); Koay et al. (2009a).

The estimation of noise standard deviation is key to accurate signal transformation. For spatially stationary noise, the noise standard derivation can be estimated from the image background via a method called probabilistic identification and estimation of noise (PIESNO) Koay et al. (2009b). For spatially non-stationary noise, a number of methods can be used Manjón et al. (2010); St-Jean et al. (2016); Manjón et al. (2013); Veraart et al. (2016). For instance, adaptive NLM (ANLM) Manjón et al. (2010) and NLSAM St-Jean et al. (2016) estimate the noise standard deviation using self-recurrent information. The local PCA method Manjón et al. (2013) introduces noise estimators for single (SIBE) or multiple (MUBE) baseline ($b = 0\,\mathrm{s/mm^2}$) images. MUBE performs PCA decomposition on multiple baseline images and then uses the noise PCA component to estimate the noise standard deviation. SIBE uses a similar strategy, but performs PCA decomposition on DW images. Using random matrix theory, the Marchenko-Pastur distribution can be used to determine an appropriate eigenvalue threshold for determining the noise PCA components Veraart et al. (2016). This method, called MPPCA Veraart et al. (2016), simultaneously estimates the threshold and the noise standard deviation. The nc-$\chi$ bias in the estimated noise standard deviation is corrected using the method described in Koay and Basser (2006). In this work, we use PIESNO and MPPCA, respectively, for stationary and non-stationary noise estimation.

### *2.2. Noise Reduction*

#### *2.2.1. x-q Space Non-Local Means*

We propose to utilize patch matching in both $x$-space and $q$-space for effective denoising. For each voxel at location $\mathbf{x}_i \in \mathbb{R}^3$, the diffusion-attenuated signal measurement $S(\mathbf{x}_i, \mathbf{q}_k)$ corresponding to wavevector $\mathbf{q}_k \in \mathbb{R}^3$ is denoised by averaging over non-local measurements that have similar $q$-patches. Note that the signal is Gaussian distributed after the transformation described in the previous section. We estimate the denoised signal as

$$\mathrm{NLM}(S)(\mathbf{x}_i, \mathbf{q}_k) = \sum_{(\mathbf{x}_j, \mathbf{q}_l) \in \mathcal{V}_{i,k}} w[i,k;j,l] S(\mathbf{x}_j, \mathbf{q}_l), \tag{1}$$

where $\mathcal{V}_{i,k}$ is the search neighborhood in $x$-$q$ space associated with $(\mathbf{x}_i, \mathbf{q}_k)$, $w[i,k;j,l]$ is the weight indicating the $q$-patch similarity between $(\mathbf{x}_i, \mathbf{q}_k)$ and $(\mathbf{x}_j, \mathbf{q}_l)$, which is determined using patch matching as described next.

Instead of restricting patch matching to $x$-space Wiest-Daesslé et al. (2007, 2008); Descoteaux et al. (2008); Yap et al. (2014), we introduce $x$-$q$ space patch matching by defining a patch in $q$-space. For each point in $x$-$q$ space, $(\mathbf{x}_i, \mathbf{q}_k)$, we define a spherical patch, $\mathcal{N}_{i,k}$, centered at $\mathbf{q}_k$ with fixed $q_k = |\mathbf{q}_k|$ and subject to a neighborhood angle $\alpha_\mathrm{p}$. The samples on this spherical patch are mapped to a disc using azimuthal equidistant projection (AEP, Section 2.2.2) before computing rotation invariant features via polar complex exponential transform (PCET, Section 2.2.3) for patch matching. Fig. 1 illustrates how patch matching is carried out in $x$-$q$ space. The search radius in $x$-space is $s$ ($2s+1$ in diameter) and the search angle in $q$-space is $\alpha_\mathrm{s}$. Matching between different shells is allowed.

In practice, $q$-space is not always sampled in a shell-like manner. To deal with this issue, we project measurement samples onto spherical patches. Each sampling point in $q$-space, $(\mathbf{x}_i, \mathbf{q}_k)$, can be seen as defining a virtual shell with radius $|\mathbf{q}_k|$. We project the signals onto this shell to form $\mathcal{N}_{i,k}$ for patch matching. This is done for example by projecting the signal measured at $(\mathbf{x}_i, \mathbf{q}_{k'})$, i.e., $S(\mathbf{x}_i, \mathbf{q}_{k'})$ to $\left(\mathbf{x}_i, \frac{|\mathbf{q}_k|}{|\mathbf{q}_{k'}|}\mathbf{q}_{k'}\right)$, taking a value that is modulated by an exponential function of difference in $b$-values, i.e.,

$$S(\mathbf{x}_i, \mathbf{q}_{k'}) \exp\left\{-\frac{(\sqrt{b_{k'}} - \sqrt{b_k})^2}{h^2_{\mathrm{projection}}}\right\}. \tag{2}$$

where $b_k = t|\mathbf{q}_k|^2$ and $b_{k'} = t|\mathbf{q}_{k'}|^2$ are the respective $b$-values and $t$ is the diffusion time. Parameter $h_{\mathrm{projection}}$ controls the attenuation of the exponential

function.

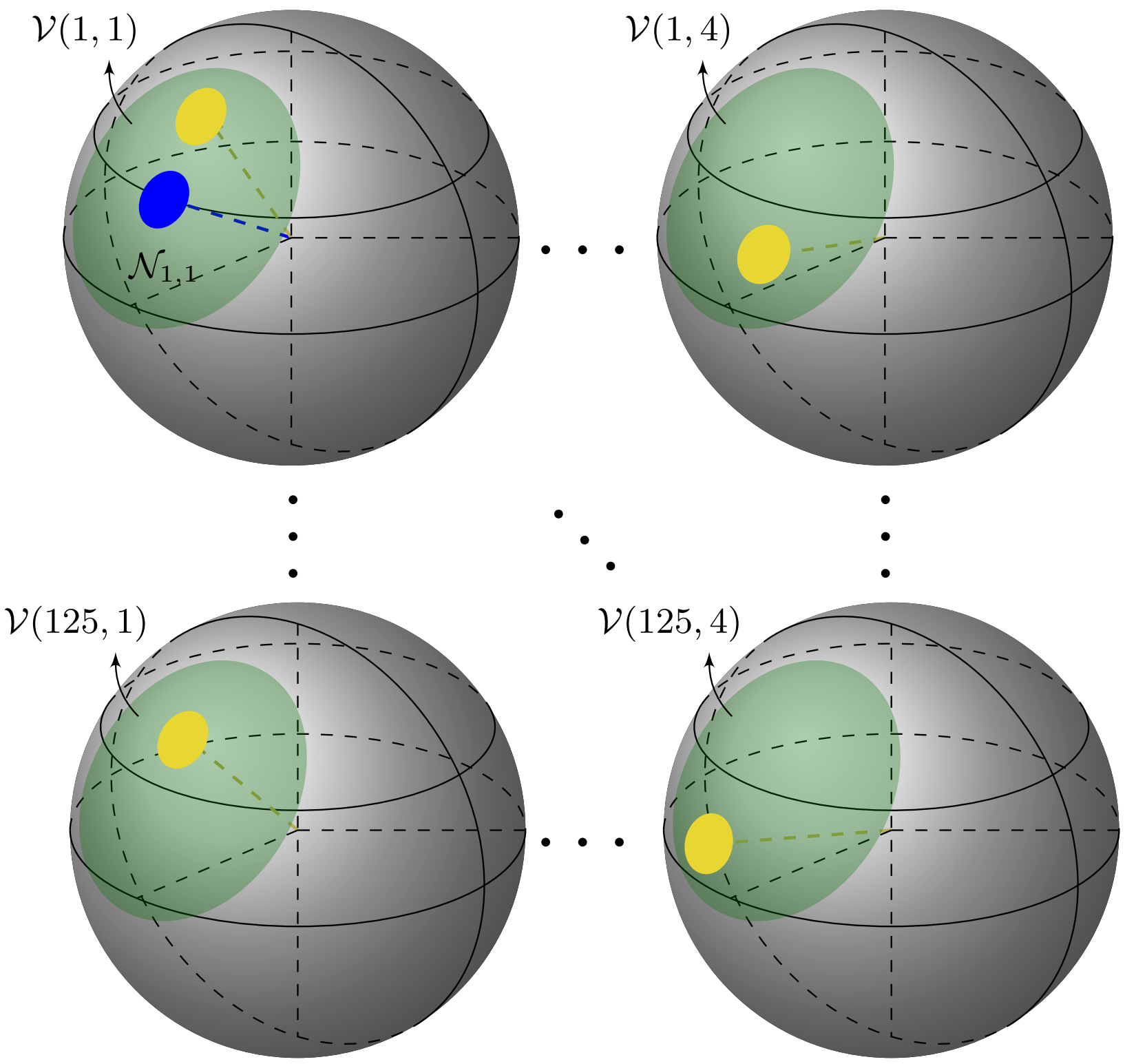

Figure 1: **Patch Matching in *x*-*q* Space.** Illustration of patch matching involving 4 shells in $q$-space and a search radius of 2 voxels (5 voxels in diameter) in $x$-space. The search neighborhood $\mathcal{V}$ is a combination of the sub-neighborhoods $\{\mathcal{V}(j,r)\}_{j=1,\ldots,5^3,r=1,\ldots,4}$ (green) associated with different locations $\{\mathbf{x}_j\}_{j=1,\ldots,5^3}$ and $b$-values $\{b_r\}_{r=1,\ldots,4}$, i.e., $\mathcal{V} = \cup_{j,r}\mathcal{V}(j,r)$, where $\mathcal{V}(j,r) \equiv \mathcal{V}(\mathbf{x}_j, b_r)$. Patch matching is carried out between the reference patch (blue) and each candidate patch (yellow) in the search neighborhood $\mathcal{V}$.

### *2.2.2. Azimuthal Equidistant Projection (AEP)*

We use AEP Wessel and Smith (2001) to map the coordinates on a sphere to a plane where the distances and azimuths of points on the sphere are preserved with respect to a reference point Wessel and Smith (2001). This provides a good basis for subsequent computation of invariant features for matching. The reference point, which in our case corresponds to the center of a spherical patch, will project to the center of a circular projection. As illustrated in Fig. 2, viewing the reference point as the 'North pole', all points along a given azimuth $\theta$ will project along a straight line from the center. In the projection plane, this line subtends an angle $\theta$ with the vertical. The distance from the center to another projected point is given as $\rho$. We represent the reference point $\hat{\mathbf{q}}_0 = \frac{\mathbf{q}_0}{|\mathbf{q}_0|}$ as spherical coordinates $(\phi_0, \lambda_0)$, with $\phi$ referring to latitude and $\lambda$ referring to longitude. We project $(\phi, \lambda)$ to a corresponding point $(\rho, \theta)$ in a 2D polar coordinate system, where $\rho$ is the radius and $\theta$ is the angle. Based on Wessel and Smith (2001), the relationship between $(\phi, \lambda)$ and $(\rho, \theta)$ is as follows:

$$\cos\rho = \sin\phi_0 \sin\phi + \cos\phi_0 \cos\phi \cos(\lambda - \lambda_0), \tag{3a}$$

$$\tan\theta = \frac{\cos\phi \sin(\lambda - \lambda_0)}{\cos\phi_0 \sin\phi - \sin\phi_0 \cos\phi \cos(\lambda - \lambda_0)}. \tag{3b}$$

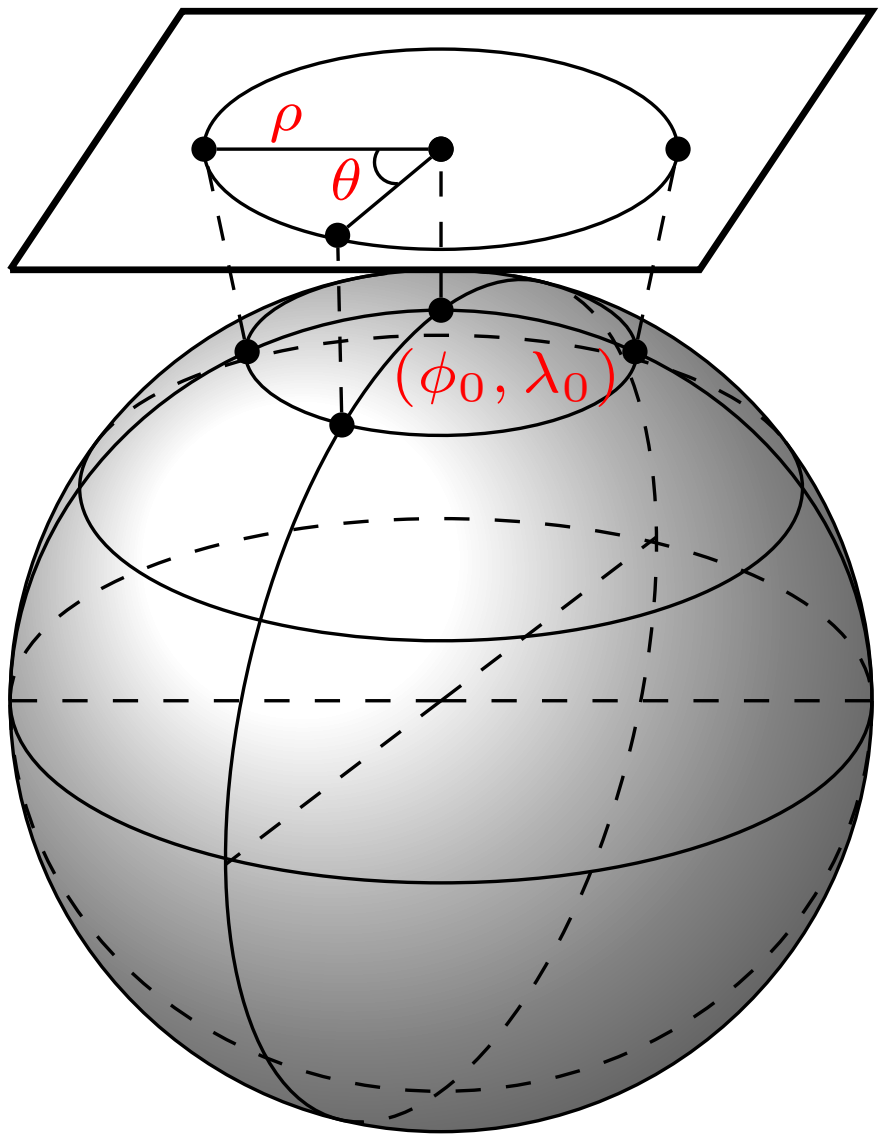

Figure 2: **Azimuthal equidistant projection (AEP).** Mapping of a spherical patch to a disc.

The projection can be described as a two-step mapping:

$$\mathbf{q} \longrightarrow (q, \phi, \lambda) \longrightarrow (q, \rho, \theta). \tag{4}$$

Note that extra care needs to be taken when using the above equations to take into consideration the fact that diffusion signals are antipodal symmetric. Prior to performing AEP, we map antipodally all the points on the sphere to the hemisphere where the reference point is located. AEP maps a $q$-space spherical patch $\mathcal{N}$ to a 2D circular patch $\widehat{\mathcal{N}}$. Note that AEP changes only the coordinates but not the actual values of the signal vector. If we let $\mathbf{S}(\mathcal{N})$ be a vector containing the values of all diffusion signals in $\mathcal{N}$, then $\mathbf{S}(\widehat{\mathcal{N}}) = \mathbf{S}(\mathcal{N})$.

### *2.2.3. Polar Complex Exponential Transform (PCET)*

After AEP, we proceed with the computation of rotation invariant features. We choose to use the polar complex exponential transform (PCET) Yap et al. (2010) for its computation efficiency and its rotation-invariance property as demonstrated in Yap et al. (2010). Rotation invariance allows matching of patches that have different orientations. Denoting an element of $\mathbf{S}(\widehat{\mathcal{N}})$ as $S(\mathbf{x}, q, \rho, \theta)$, the PCET of order $n$, $|n|= 0, 1, 2, \ldots, \infty$, and repetition $l$, $|l|= 0, 1, 2, \ldots, \infty$, is defined as

$$M_{n,l}(\widehat{\mathcal{N}}) = \frac{1}{\pi} \int_{(\mathbf{x},q,\rho,\theta)\in\widehat{\mathcal{N}}} [H_{n,l}(\rho, \theta)]^* S(\mathbf{x}, q, \rho, \theta) \rho \,\mathrm{d}\,\rho \,\mathrm{d}\,\theta, \tag{5}$$

where $[\cdot]^*$ denotes the complex conjugate and $H_{n,l}(\rho, \theta)$ is the basis function defined as $H_{n,l}(\rho, \theta) = e^{i2\pi n\rho^2} e^{il\theta}$. It can be easily verified that $|M_{n,l}(\widehat{\mathcal{N}})|$ is invariant to rotation Yap et al. (2010). Interested readers are referred to Yap et al. (2010) for mathematical and implementation details. $|M_{n,l}(\widehat{\mathcal{N}})|$'s up to maximum order $m$, i.e., $-m \leq l, n \leq m$, are concatenated into a feature vector $\mathbf{M}(\widehat{\mathcal{N}})$.

### *2.2.4. Patch Matching*

Let $\mathbf{M}(\widehat{\mathcal{N}}_{i,k})$ be the feature vector of the projected patch $\widehat{\mathcal{N}}_{i,k}$, the matching weight $w[i, k; j, l]$ is defined as

$$w[i, k; j, l] = \frac{1}{Z_{i,k}} w_{\mathbf{M}}[i, k; j, l] w_{\mathrm{b}}[i, k; j, l], \tag{6}$$

with

$$w_{\mathbf{M}}[i, k; j, l] = \exp\left\{ -\frac{\|\mathbf{M}(\widehat{\mathcal{N}}_{i,k}) - \mathbf{M}(\widehat{\mathcal{N}}_{j,l})\|_2^2}{h_{\mathbf{M}}^2(i, k)} \right\}, \tag{7}$$

$$w_{\mathrm{b}}[i, k; j, l] = \exp\left\{ -\frac{(\sqrt{b_k} - \sqrt{b_l})^2}{h_{\mathrm{b}}^2} \right\}, \tag{8}$$

where $Z_{i,k}$ is a normalization constant to ensure that the weights sum to one:

$$Z_{i,k} = \sum_{(\mathbf{x}_j, \mathbf{q}_l) \in \mathcal{V}_{i,k}} w_{\mathbf{M}}[i, k; j, l] w_{\mathrm{b}}[i, k; j, l]. \tag{9}$$

Here $h_{\mathbf{M}}(i, k)$ is a parameter controlling the attenuation of the first exponential function. As in Coupé et al. (2008), we set $h_{\mathbf{M}}(i, k) = \sqrt{2\beta_{\mathbf{M}}\hat{\sigma}_{i,k}^2 |\mathbf{M}(\widehat{\mathcal{N}}_{i,k})|}$, where $\beta_{\mathbf{M}}$ is a constant Coupé et al. (2008), $\hat{\sigma}_{i,k}$ is the noise standard deviation, which is computed spatial-adaptively Veraart et al. (2016). Similarly, $h_{\mathrm{b}} = \sqrt{2}\sigma_{\mathrm{b}}$ controls the attenuation of the other exponential function, where $\sigma_{\mathrm{b}}$ is a scale parameter.

## 3. Experiments

### *3.1. Datasets*

#### *3.1.1. Synthetic Data*

For quantitative evaluation, a synthetic multi-shell dataset was generated using Phantom$\alpha$s Caruyer et al. (2014), which is a toolbox for simulation of DMRI data with complex fiber geometries. We use the geometric model designed for ISBI 2013 HARDI challenge[1], which consists of various configurations such as branching, crossing, and kissing. The parameters used for data simulation were chosen to be consistent with the real data described in Section 3.1.3: $b = 1000, 2000, 3000\,\mathrm{s/mm}^2$, 90 gradient directions per shell, $55 \times 55$ voxels with resolution $2 \times 2\,\mathrm{mm}^2$.

Stationary and non-stationary nc-$\chi$ noise with 1, 4, and 8 channels and level 5%, 7.5%, 10% was added to the data. For an $N$-channel receiver coil, the measured signal $Y_N$ with nc-$\chi$ noise is given by Constantinides et al. (1997); Koay and Basser (2006); Koay et al. (2009a)

$$Y_N = \sqrt{\sum_{k=1}^{N} [(\mu_{\mathrm{R}}(k) + X_{\mathrm{R}}(k))^2 + (\mu_{\mathrm{I}}(k) + X_{\mathrm{I}}(k))^2]}, \tag{10}$$

where $\mu_{\mathrm{R}}(k)$ and $\mu_{\mathrm{I}}(k)$ are respectively the real and imaginary parts of the true complex signal from the $k$-th receiver coil. $X_{\mathrm{R}}(k)$ and $X_{\mathrm{I}}(k)$ are two random variables that follow the same Gaussian distribution with standard deviation $\gamma\sigma$: $X_{\mathrm{R}}(k) \sim \mathcal{N}(0, \gamma\sigma)$ and $X_{\mathrm{I}}(k) \sim \mathcal{N}(0, \gamma\sigma)$. In the absence of

[1] http://hardi.epfl.ch/static/events/2013_ISBI/

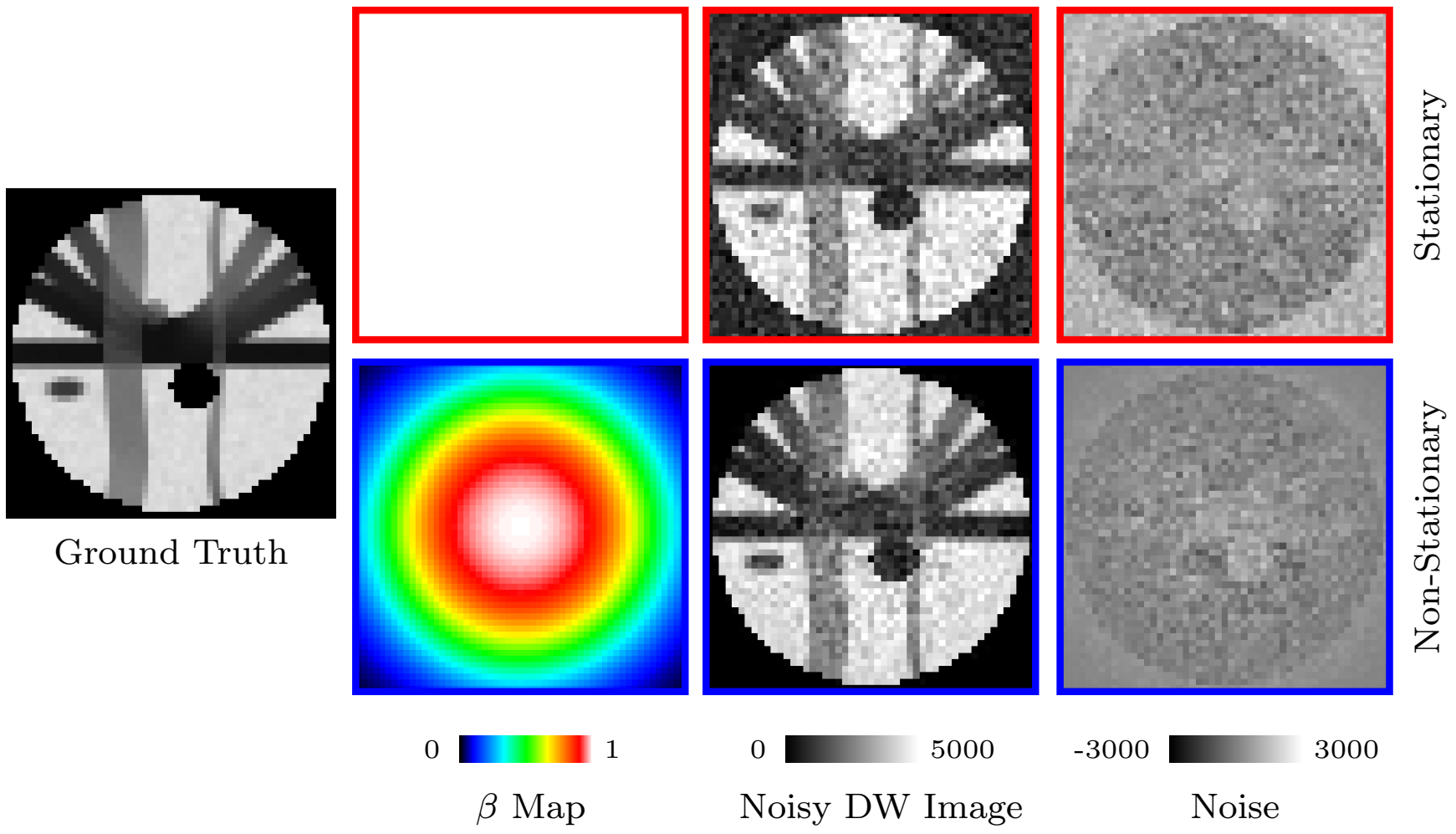

Figure 3: **Noise Simulation.** 5% 4-channel nc-$\chi$ noise with spatially constant and varying $\gamma$-maps. DW images with $b = 1000\,\mathrm{s/mm^2}$ are shown here.

noise, the true signal $\mu_N$ can be expressed as

$$\mu_N = \sqrt{\sum_{k=1}^{N} [\mu_{\mathrm{R}}^2(k) + \mu_{\mathrm{I}}^2(k)]}. \tag{11}$$

For stationary noise, we set $\gamma = 1$. For non-stationary noise, $\gamma$ varies spatially. We set $\sigma$ as $p$ percent of the maximum data intensity value $v$, i.e., $\sigma = v(p/100)$. Examples of stationary and non-stationary $\gamma$-maps used in this work are shown in Fig. 3. Example images for different numbers of channels are shown in Fig. 4.

### *3.1.2. Repeated Acquisition Data*

We acquired the brain DMRI data of an adult 25 times using a Siemens 3T Magnetom Prisma MR scanner with the following imaging protocol: $b = 3000\,\mathrm{s/mm^2}$, 42 gradient directions, $140 \times 140$ imaging matrix, voxel size $1.5\times1.5\times1.5\,\mathrm{mm^3}$, TE=89 ms, TR=2,513 ms, 32 receiver coils. We performed signal transformation Eichner et al. (2015) and eddy correction Andersson and Sotiropoulos (2016) for each dataset. The 25 processed datasets were averaged to form a gold standard with improved SNR for evaluation purposes. Informed written consent was obtained from the subject and the experimental protocols were approved by the Institutional Review Board of the University of North Carolina (UNC) School of Medicine. The study was carried out in accordance with the approved guidelines.

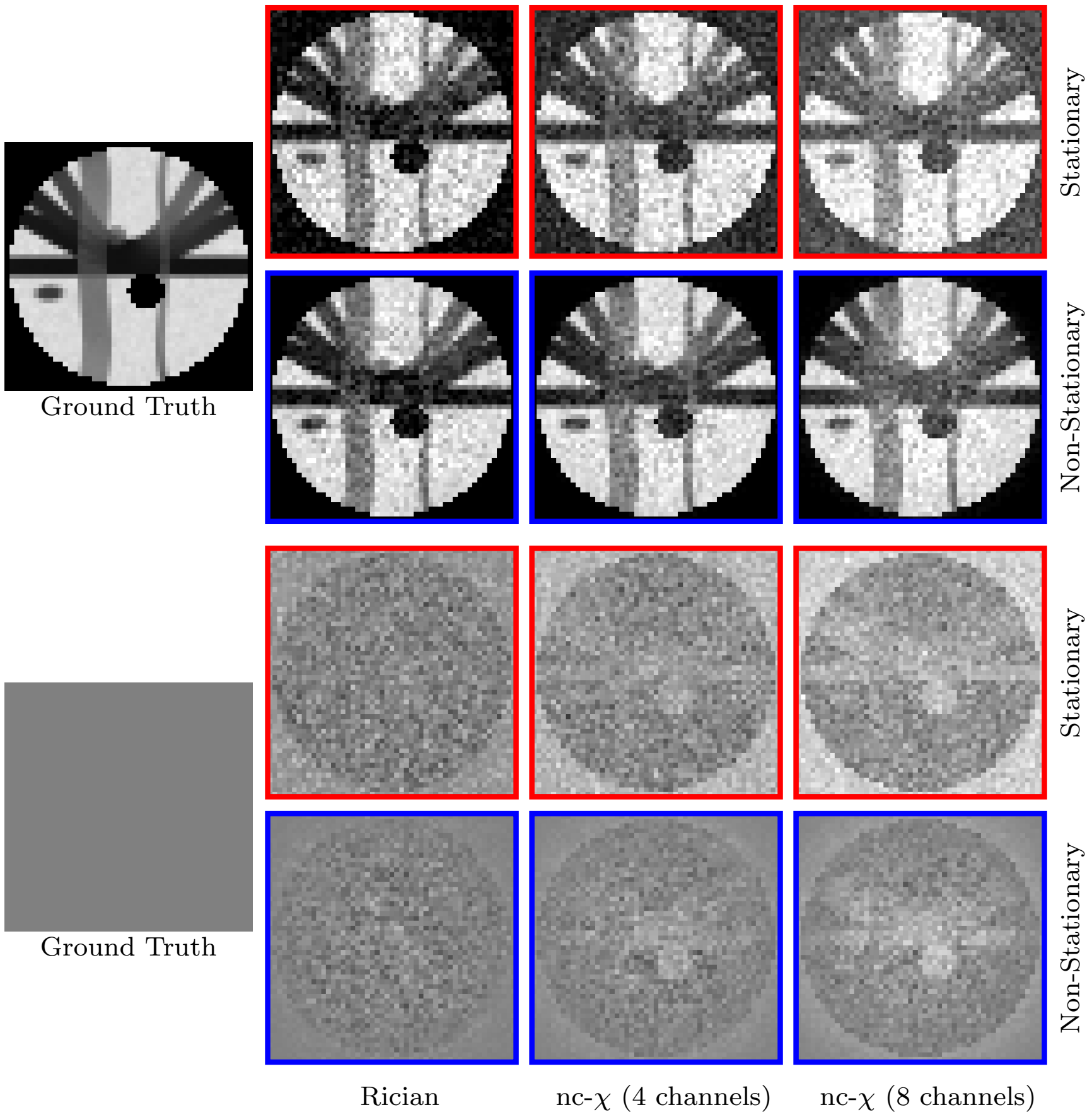

Figure 4: **Synthetic Data.** Some examples of the synthetic data ($b = 1000\,\mathrm{s/mm^2}$) for stationary and non-stationary noise with different numbers of channels.

#### 3.1.3. Multi-Shell HCP Data

The diffusion dataset of one subject randomly selected from the Human Connectome Project (HCP) Van Essen et al. (2013) was used for evaluation. Instead of the minimally preprocessed data, we used the unprocessed data to avoid alteration of the noise distribution Veraart et al. (2013). A customized Siemens 3T Connectome Skyra housed at Washington University in St. Louis was used for scanning. The imaging protocol was as follows: $145 \times 174$ imaging matrix, $1.25 \times 1.25 \times 1.25$ mm$^3$ resolution, TE=89 ms, TR=5,500 ms, 32-channel receiver coil. CMS reconstruction was performed using SENSE1 Sotiropoulos et al. (2013), resulting in non-stationary Rician noise distribution.

### 3.2. Experimental Setting

#### 3.2.1. Parameter Settings

For all experiments, the following parameters were used for $x$-$q$ space non-local means denoising (XQ-NLM):

1. The maximum order of PCET was set to $m = 4$, which we found to be sufficient for patch characterization.
2. Following Coupé et al. (2008), we set $s = 2$ voxels. Instead of $\beta_{\mathbf{M}} = 1$ as suggested in Coupé et al. (2008), we set $\beta_{\mathbf{M}} = 0.1$ since we have a greater number of patch candidates by considering the joint $x$-$q$ space. Based on the theory of kernel regression Silverman (1998), reducing the bandwidth when the sample size is large reduces bias.
3. The smallest non-zero value for $|\sqrt{b_k} - \sqrt{b_l}|$ is around 10 (i.e., $\sqrt{3000} - \sqrt{2000} \approx 10$). We set $\sigma_{\mathrm{b}} = 10/2 = 5$.

4. Since we were using shell-sampled data in our evaluations, we set $h_{\text{projection}}$ to a small value (0.1) to disable projection.
5. In our case, the minimal angular separation of the gradient directions is around 15° for each shell. We set the $q$-space patch angle and $q$-space search angle to twice this value, i.e., $\alpha_{\text{p}} = \alpha_{\text{s}} = 2 \times 15° = 30°$.

### 3.2.2. Methods for Comparison

We compared XQ-NLM with the following methods:

1. **Adaptive non-local means (ANLM):** ANLM Manjón et al. (2010) is an extension of the NLM algorithm which removes spatially non-stationary noise. Based on Manjón et al. (2010), we set the patch radius to 1 and search radius to 2.
2. **Non-local spatial and angular matching (NLSAM):** NLSAM St-Jean et al. (2016) consists of three major steps, i.e., (i) Signal transformation so that the signals are Gaussian distributed; (ii) 4D block construction by considering diffusion-weighted images within an angular neighborhood; (iii) Noise removal using sparse representation. Based on St-Jean et al. (2016), we set the patch radius to 1 and use 5 angular neighbours.
3. **Marchenko-Pastur principle component analysis (MPPCA):** By observing the fact that noise-only egienvalues follow a Marchenko-Pastur distribution, MPPCA Veraart et al. (2016) determines the threshold for PCA denoising automatically. Based on Veraart et al. (2016), we set the window size of MPPCA to $5 \times 5 \times 5$.

For fair comparison, we used the non-stationary noise field estimated by MPPCA for ANLM, NLSAM, and XQ-NLM. For stationary noise, the noise standard deviation was determined using PIESNO Koay et al. (2009b).

#### *3.2.3. Evaluation Methods*

Quantitative and qualitative evaluations were performed in our experiments.

1. **Peak-to-signal-ratio (PNSR):** We used PSNR as the metric for performance evaluation. PSNR is defined as

$$\mathrm{PSNR} = 20 \log_{10} \frac{\mathrm{MAX}}{\mathrm{RMSE}}, \tag{12}$$

   where RMSE is the root mean square error computed between the denoised image and the ground truth noise free image in the brain region; MAX is the maximum signal value.

2. **RMSE map:** Pixelwise accuracy was evaluated using the RMSE computed between the denoised signal vector at each voxel location with respect to the ground truth.
3. **FA images:** We computed the FA image using the weighted linear tensor fitting Basser et al. (1994) implemented in Camino Cook et al. (2006).
4. **Mean absolute difference (MAD):** We computed the absolute difference (AD) map between each FA image and the gold standard. MAD is computed either across voxels or across repetitions.
5. **ODFs:** We further evaluate the influence of denoising on fiber ODF estimates. Based on the method presented in Yap et al. (2016), we computed the fiber ODFs and visually inspected their quality.

6. **Tract bundles:** Based on the estimated fiber ODFs, we used a multi-directional streamline algorithm Mori et al. (1999); Stieltjes et al. (2001) for whole brain tractography. The parameters we used are as follows: The number of ODF peaks detected per voxel is restricted to 3, the voxels with FA values larger than 0.4 are selected as seeds, the stopping FA value is set to 0.2, and the maximum allowed turning angle is set to $60^{\circ}$. Following Wakana et al. (2007), we extracted each tract bundle of interest by requiring the tracts in the bundle to traverse a set of regions of interest (ROIs).

### *3.3. Results*

#### *3.3.1. Patch Matching*

We first evaluated the performance of the proposed $q$-space patch matching. The results, shown in Fig. 5, indicate that the new patch matching scheme is robust to the variation of local fiber orientations. This allows XQ-NLM to use information from differentially oriented signal profiles for effective denoising.

#### *3.3.2. Synthetic Data*

The PSNR results shown in Figs. 6 and 7 indicate that all methods improve the PSNR, but XQ-NLM performs best for all noise levels. MPPCA outperforms NLSAM when the number of coils is small. With the increase of noise level and number of coils, the performance of MPPCA drops dramatically and performs worse than NLSAM. Compared with NLSAM, the next best method, the largest improvement given by XQ-NLM is 6.42 dB in the case of 10% stationary Rician noise.

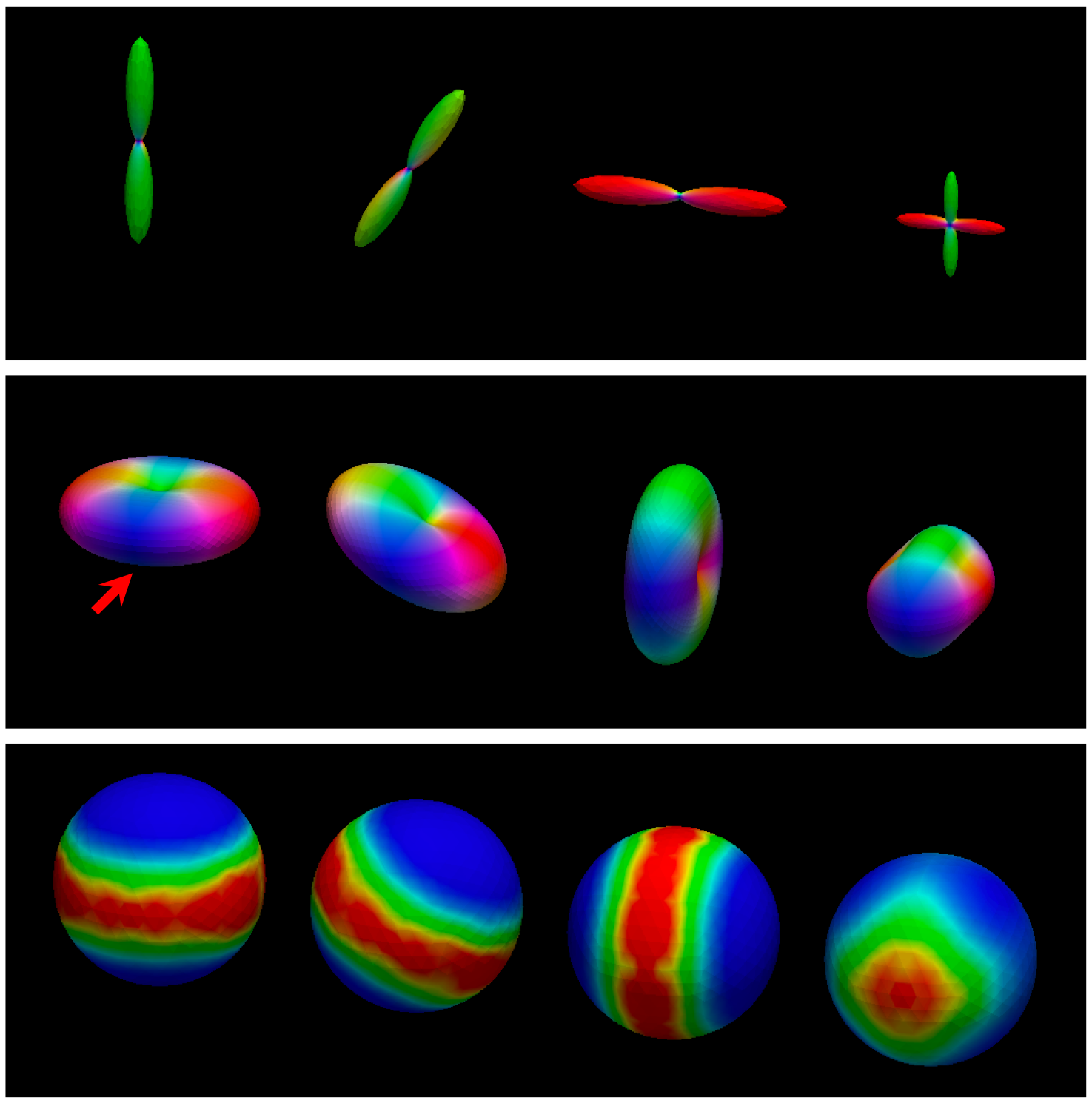

Figure 5: ***q*-Space Patch Matching.** Fiber ODFs are shown in the top row for reference. The middle row shows the profiles of the diffusion signals. Patch matching is performed using the point marked by the red arrow as the reference. The bottom row shows the matching results of signal profiles in different orientations. Warm colors indicate greater agreement, cool colors indicate otherwise. The $b$-value used for data simulation equals 1000 s/mm$^2$. For a better visualization of 3D glyphs, we use a slightly oblique view rather than a typical horizontal view.

The denoised DW images, shown in Fig. 8, indicate that XQ-NLM is able to preserve sharp edges and effectively remove noise, thanks to the robust $q$-space patch matching mechanism, as demonstrated in Fig. 5. In contrast, ANLM and NLSAM blur edges and MPPCA is unable to sufficiently remove noise.

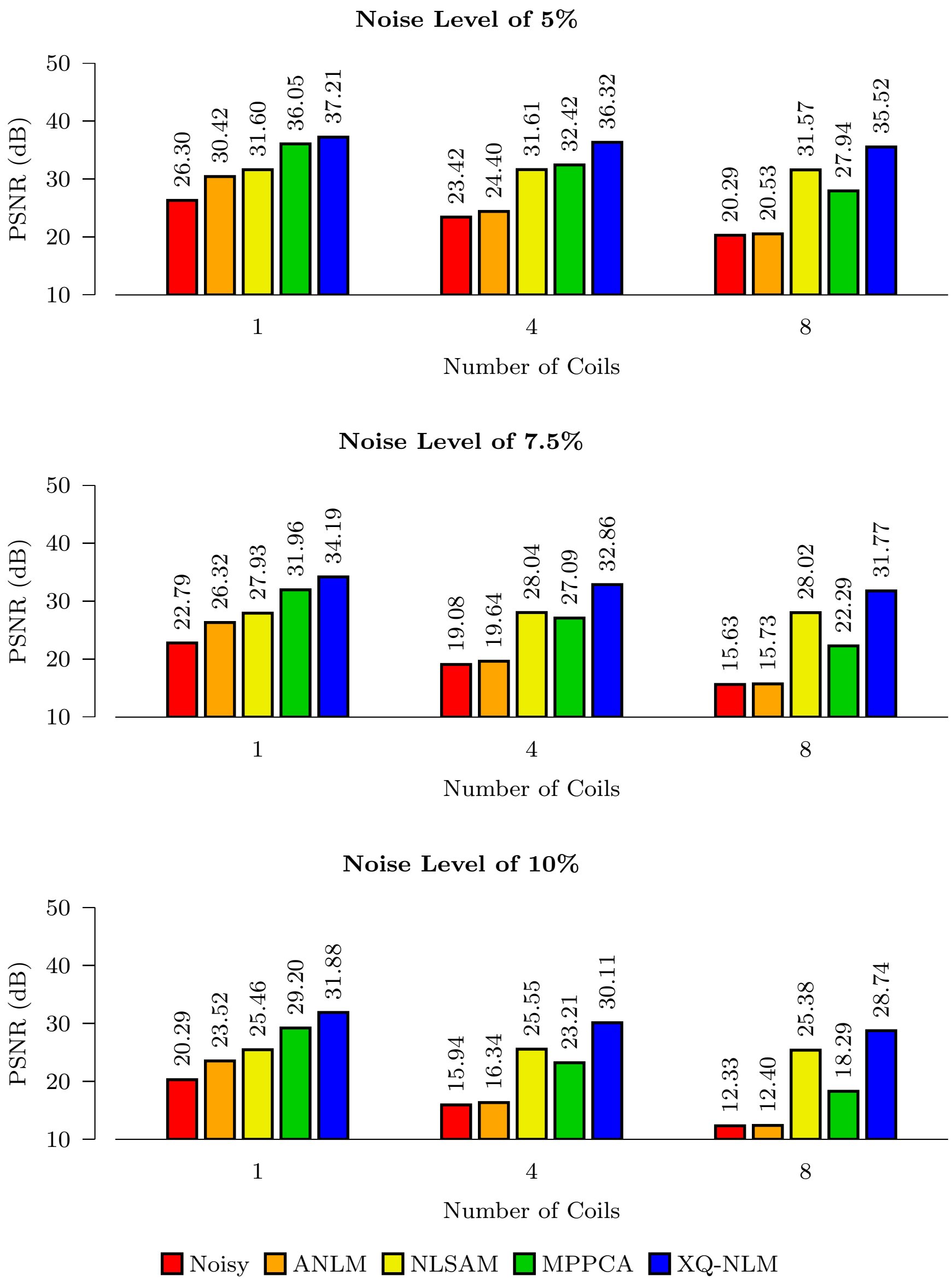

Figure 6: **PSNR Comparison – Stationary Noise.** Quantitative evaluation of denoising performance using synthetic data with spatially stationary noise.

For better comparison, we computed the RMSE map of a denoised dataset with respect to the ground truth. The results, shown in Fig. 9, indicate that

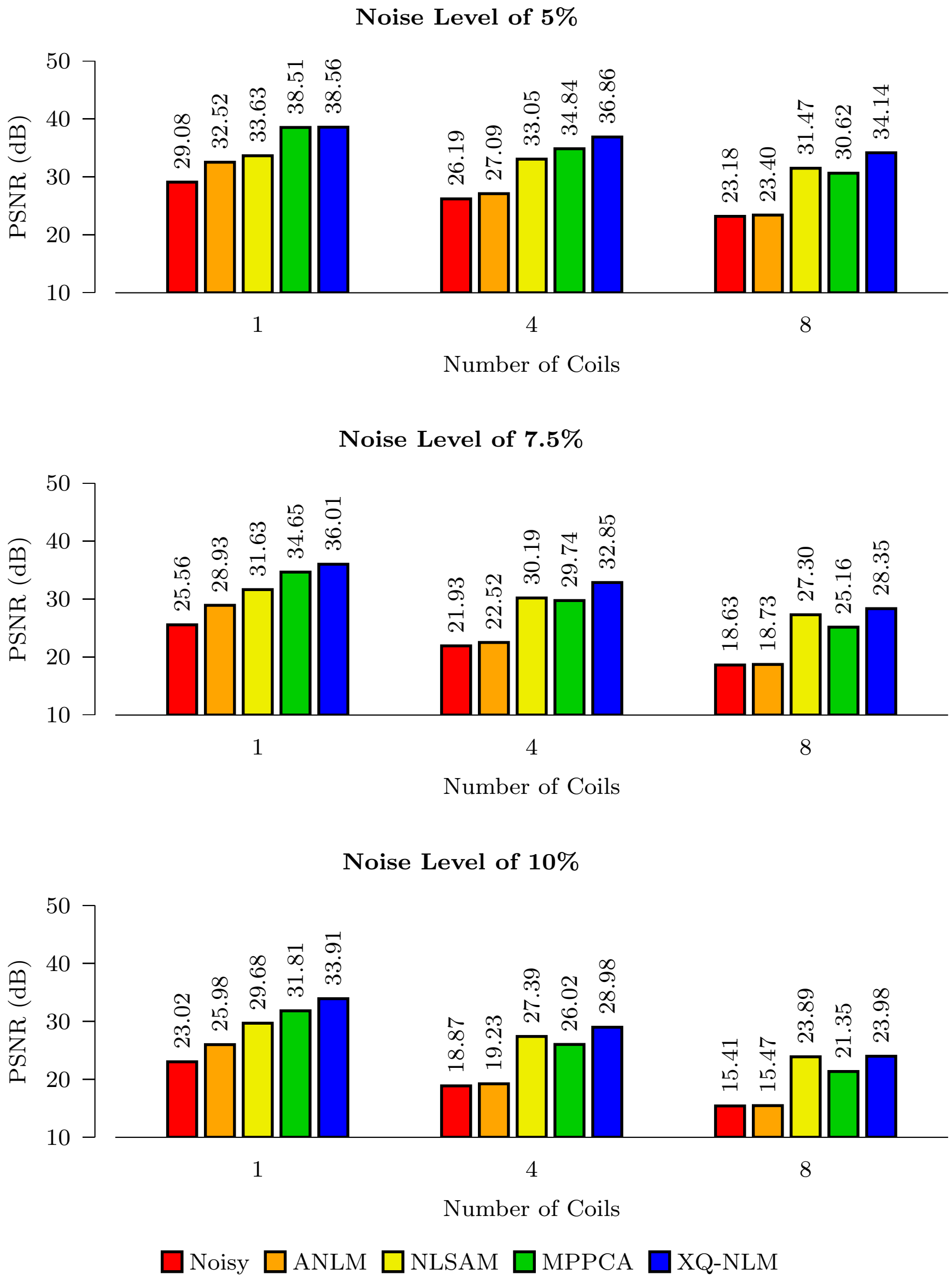

Figure 7: **PSNR Comparison – Non-Stationary Noise.** Quantitative evaluation of denoising performance using synthetic data with spatially non-stationary noise.

XQ-NLM significantly reduces the RMSE. The improvement is especially apparent at the boundaries, compared with ANLM. NLSAM slightly improves the results, but is still problematic at boundaries. MPPCA performs better than ANLM and NLSAM for edges, but fails to remove noise sufficiently. Overall, XQ-NLM gives superior performance with edge-preserving denoising performance.

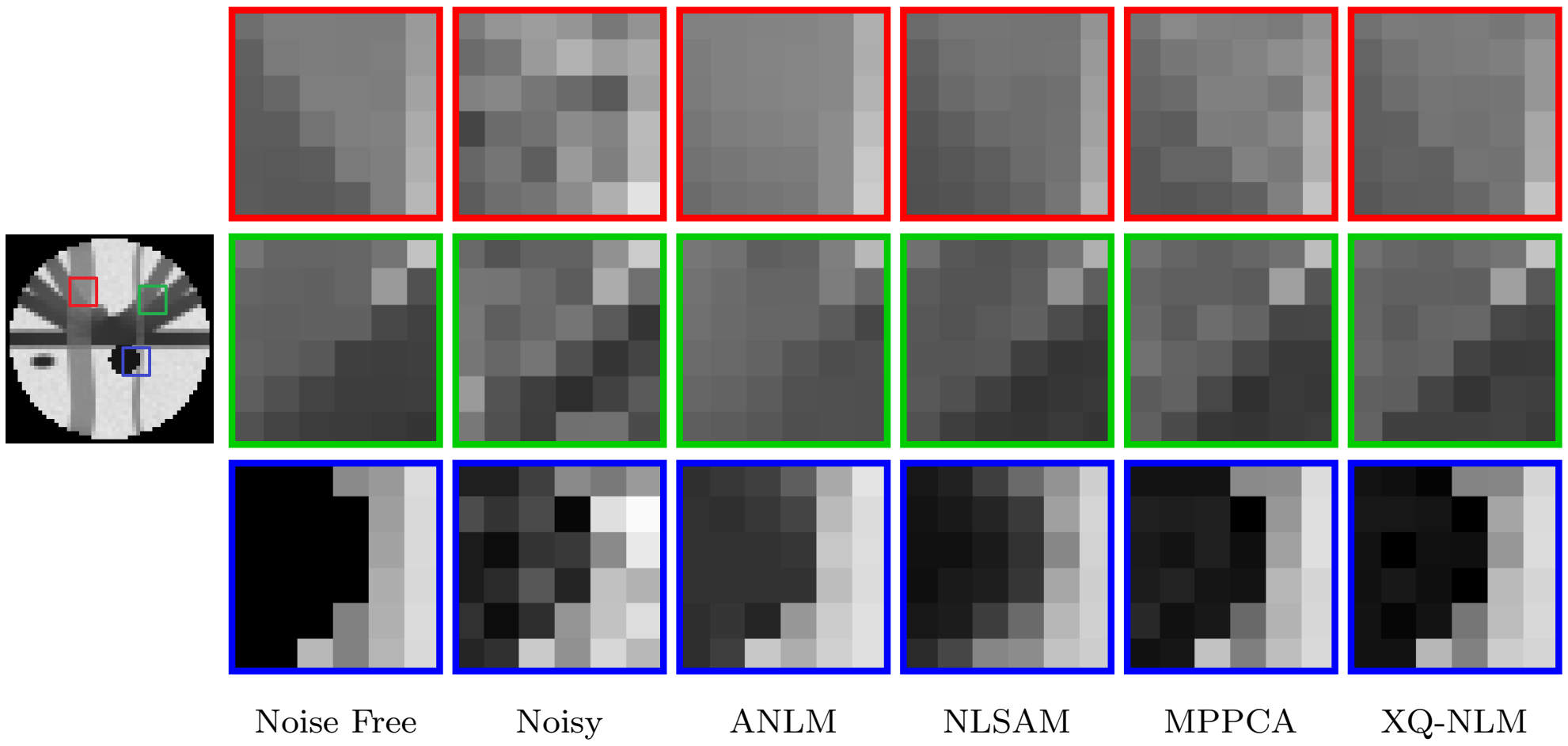

Figure 8: **DW Images – Synthetic Data.** The synthetic data ($b = 1000\,\mathrm{s/mm^2}$) with 5% 4-channel spatially non-stationary nc-$\chi$ noise was used in the evaluation.

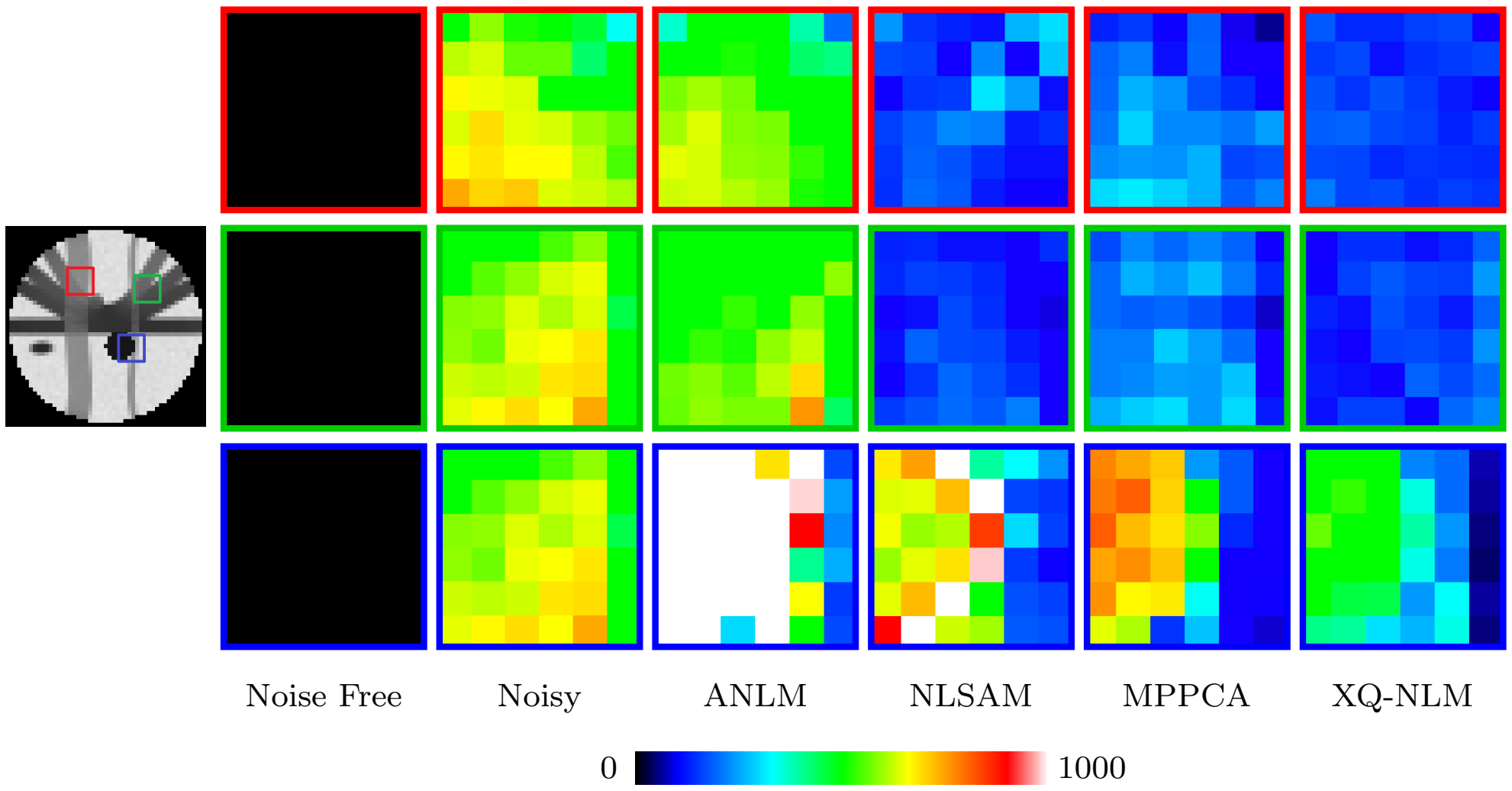

Figure 9: **RMSE Maps.** Similar to Fig. 8, but showing RMSE maps. The maximum value 1000 in the color bar is 10% of the maximum intensity value.

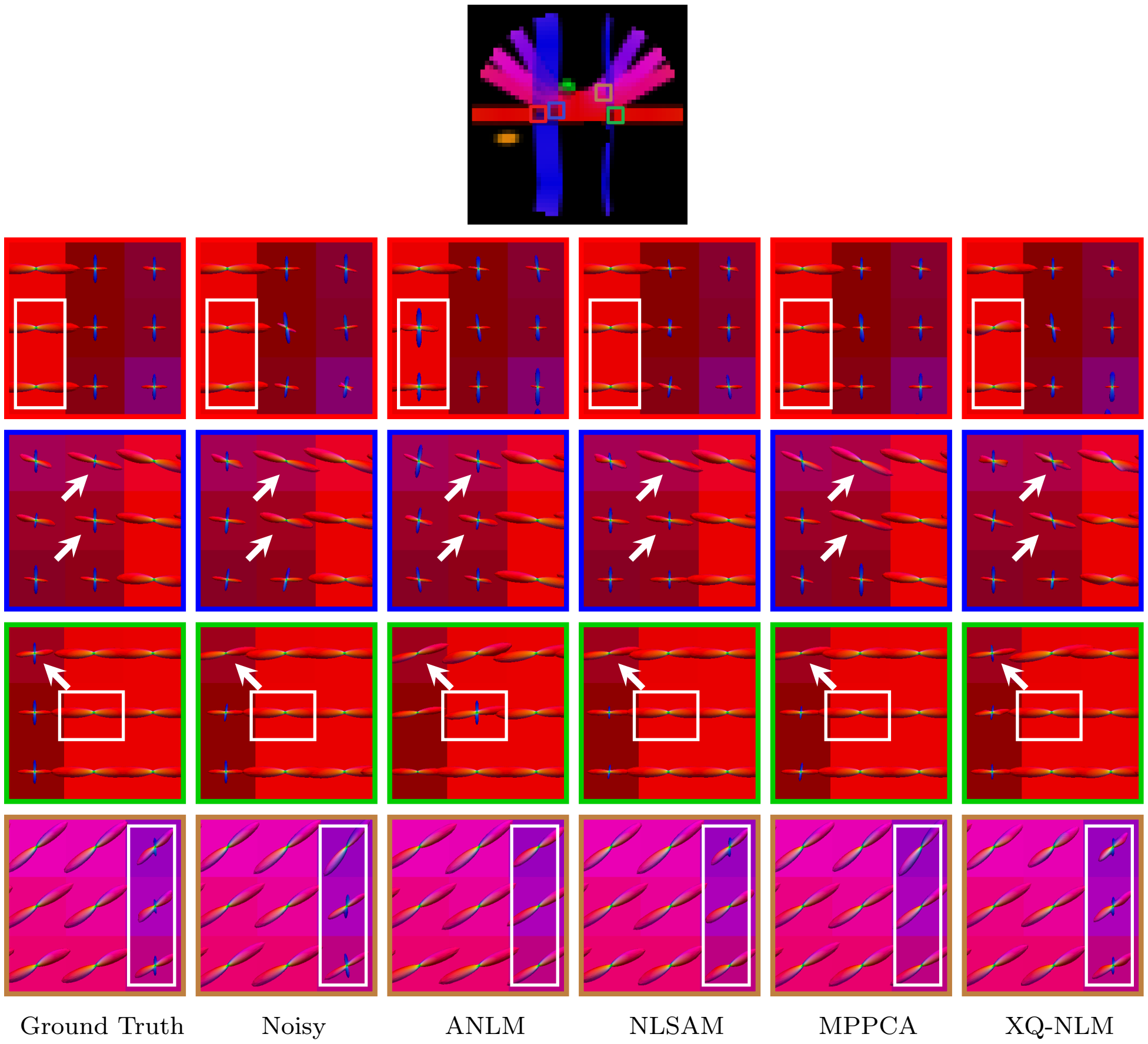

Figure 10: **Fiber ODFs – Synthetic Data.** White matter fiber ODFs for synthetic data.

The ODFs, shown in Fig. 10, indicate that XQ-NLM gives results that are very close to the ground truth. In contrast, ANLM, NLSAM and MPPCA lead to incorrect results, as marked by the white arrows. When the smoothing effect is strong, as in ANLM, spurious peaks are more likely to occur at the boundaries. From the results given by ANLM in Fig. 10, we can observe that spurious peaks are introduced in the single direction ODFs marked by the white rectangle. These incorrect peaks are introduced from the neighboring two-direction ODFs due to boundary smoothing, as can be observed from Fig. 8.

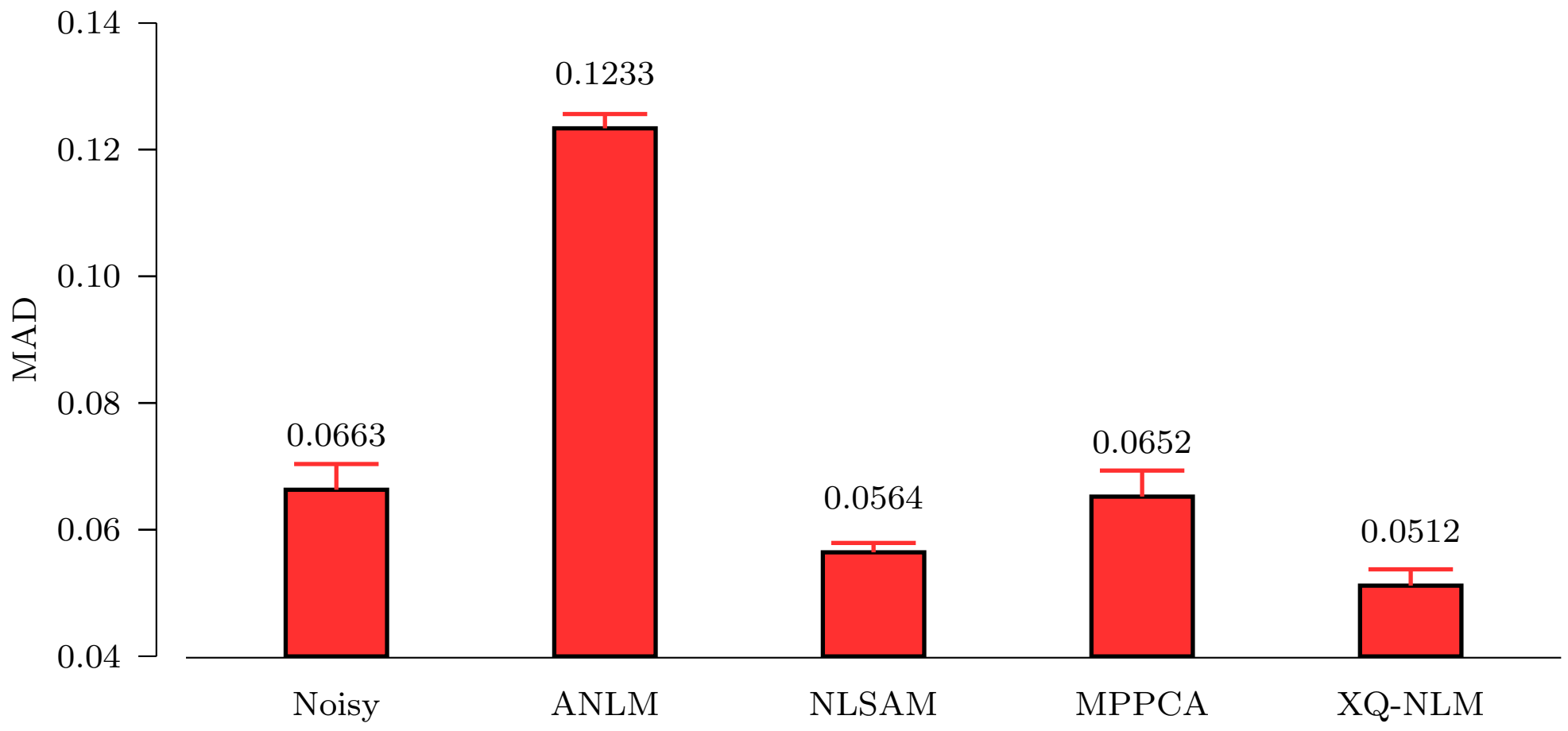

Figure 11: **MAD Comparison of FA images.** For each dataset, we computed the MAD values for the FA images given by the different methods. The means and standard deviations are shown.

### *3.3.3. Repeated Acquisition Data*

Fig. 11 indicates that XQ-NLM gives the lowest FA mean MAD values, computed over the 25 datasets, with respect to the gold standard. Fig. 12 shows the FA images of one randomly selected dataset. XQ-NLM gives a sharper FA image with preserved details even in the cortical regions. We further computed the FA MAD map averaged across 25 repetitions. The results, shown in Fig. 13, indicate that XQ-NLM gives on overall the lowest

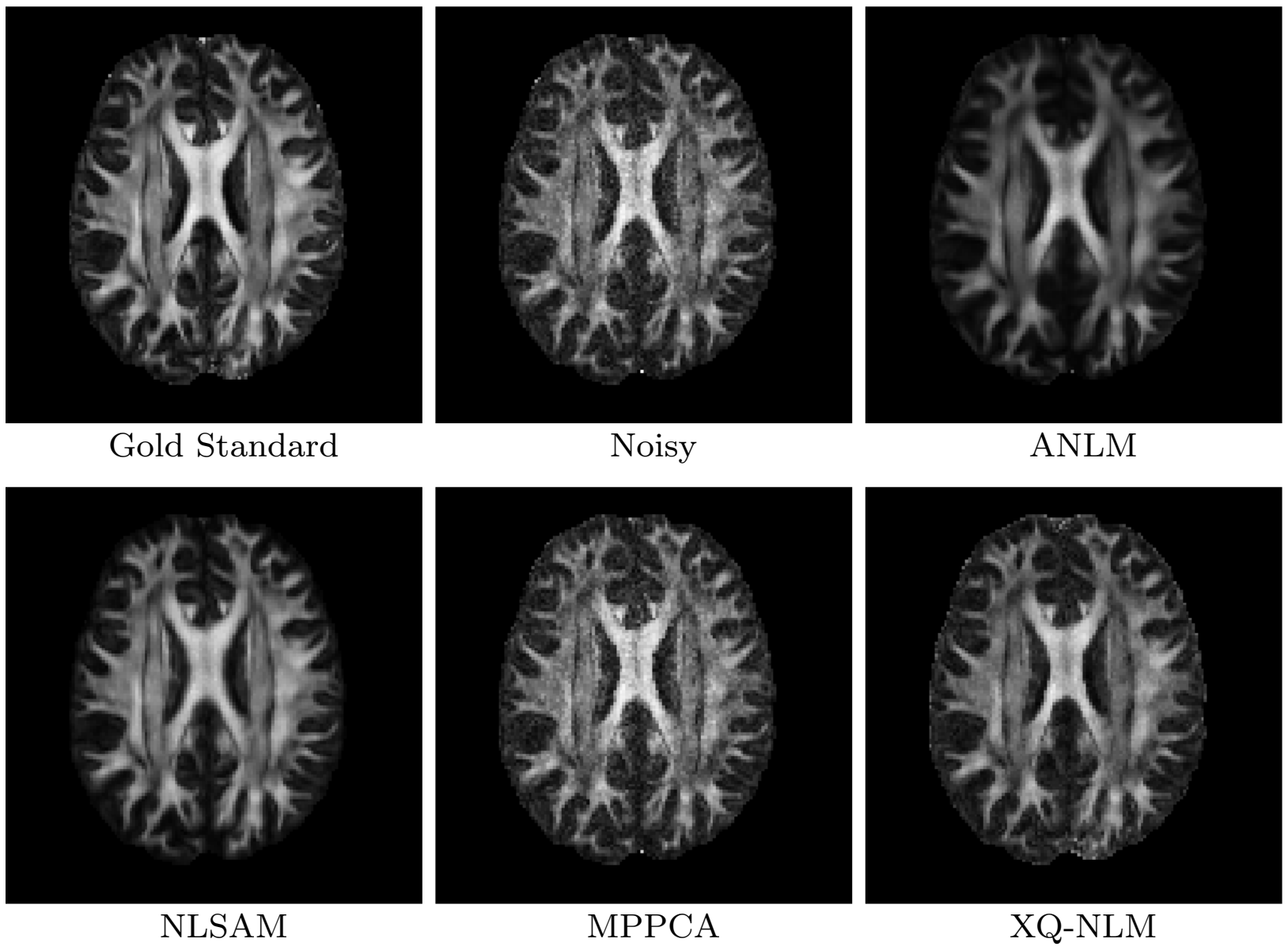

Figure 12: **FA Images.** FA images of one randomly selected dataset.

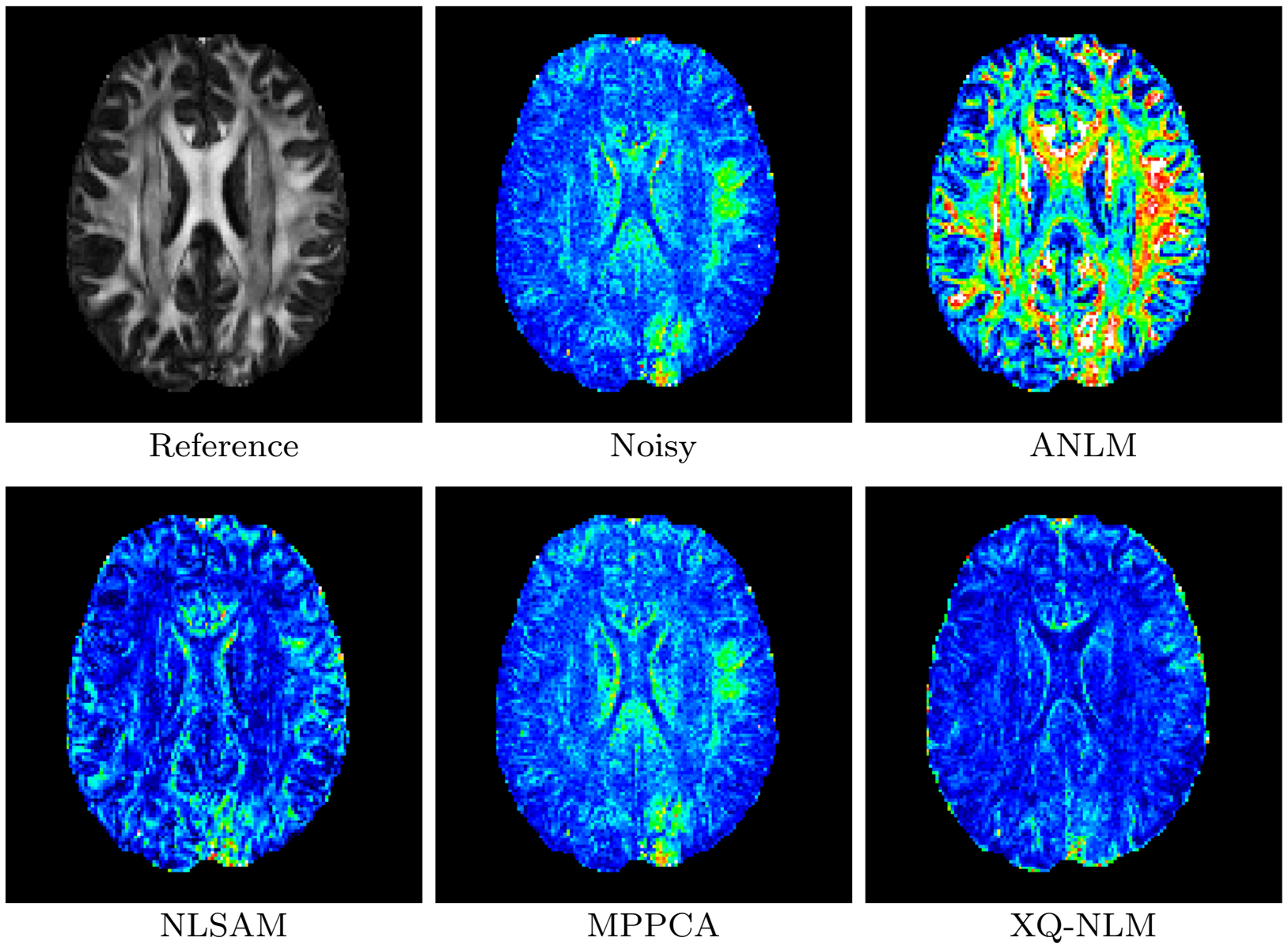

Figure 13: **FA MAD Maps.** Gold standard FA image and the MAD maps given by the different methods.

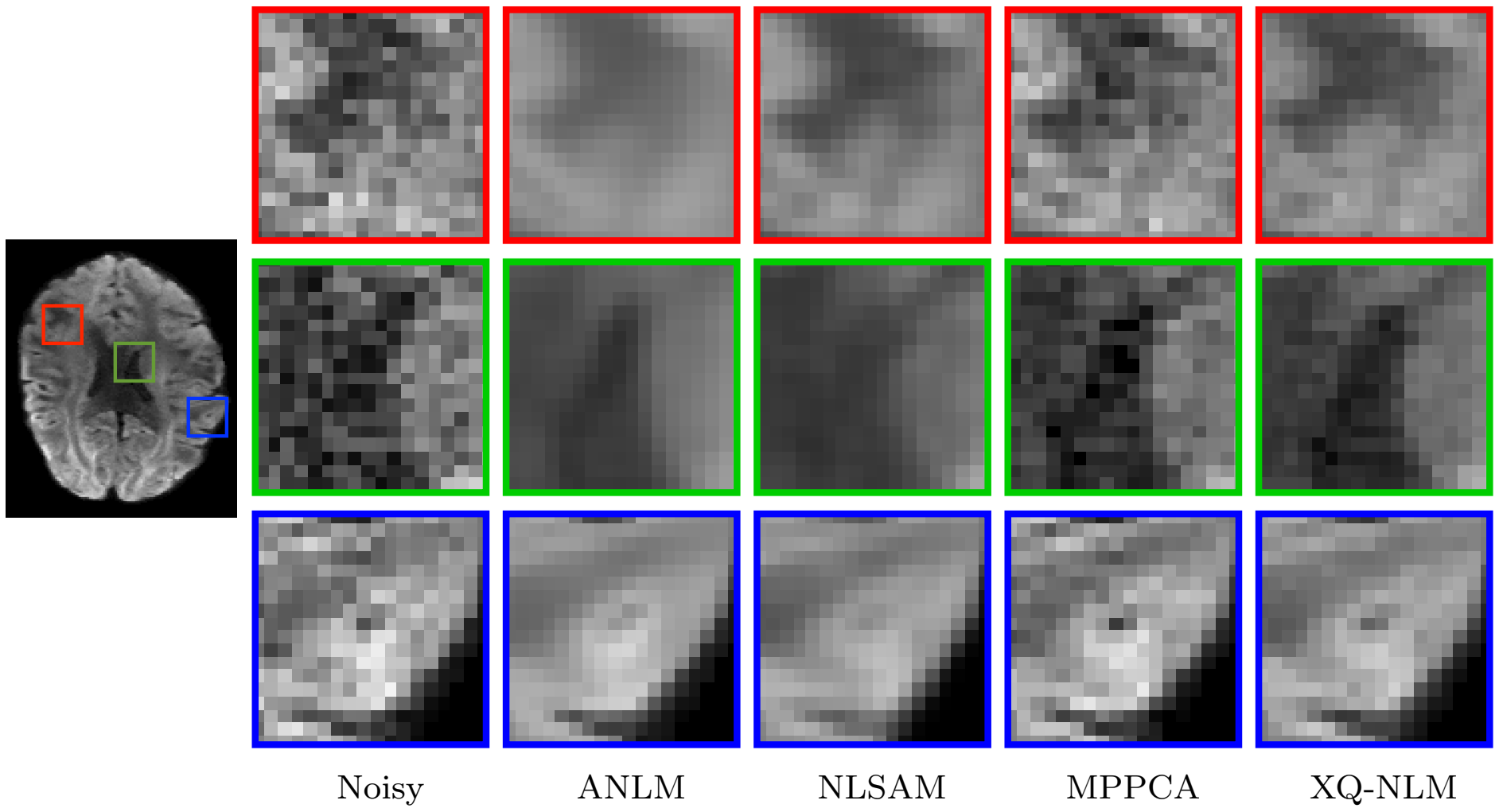

Figure 14: **DW Images – HCP Data.** DW images of the real data ($b = 1000\,\mathrm{s/mm^2}$) denoised by various methods.

MAD values, further confirming the advantages of XQ-NLM. The superior performance of XQ-NLN can be attributed to the fact that XQ-NLM is able to preserve edges while effectively remove noise. Similar to the observations based on the synthetic data experiments, ANLM and NLSAM over-smooth the data and here result in lower FA values. The synthetic data experiments also show that MPPCA does not completely remove noise. Here, this results in noisy FA images and a MAD map that resembles that of the noisy data.

### *3.3.4. Multi-Shell HCP Data*

The results shown in Fig. 14 indicate that XQ-NLM yields markedly improved edge-preserving results in the cortical regions compared with ANLM, NLSAM, and MPPCA, especially at the boundaries.

## 4. Discussion

### *4.1. Factors Contributing to the Effectiveness of XQ-NLM*

XQ-NLM demonstrates superb denoising and edge-preserving performance, which can be attributed mainly to the following factors:

- Patch matching in $q$-space allows information from curved structures to be used for denoising. This dramatically increases the effective sample size and improves the chances of finding matching information.

- NLM can be seen as kernel regression in patch space Yap et al. (2014) and large kernel bandwidths are known to introduce bias Silverman (1998). The increase in sample size allows us to utilize a tighter matching criterion, i.e., smaller kernel bandwidths, so that estimation bias can be reduced.

- The diffusion signal profile captured in each voxel is in general smooth with less abrupt changes. This again improves effective sample size because sharp changes generally imply structural peculiarity and hence greater challenges in finding matching information.

- Diffusion signal profiles have simpler shapes. This implies greater recurrence in signal patterns and hence more effective NLM denoising with lesser artifacts caused by the *rare patch effect* Duval et al. (2011); Deledalle et al. (2012); Salmon and Strozecki (2012).

### *4.2. Mitigating Noise-Induced Bias*

Unlike Gaussian noise, the signal dependency of nc-$\chi$ noise complicates the analysis of the CMS. The noise floor resulting from nc-$\chi$ noise leads to biased diffusion model fits and inaccurate signal averaging. This can be avoided by employing post-processing signal transformation techniques, such as the one described in Section 2.1.2. Alternatively, if phase images are available, real-valued diffusion data can be extracted based on the approach described in Eichner et al. (2015). Essentially, the method eliminates shot-to-shot phase variations of complex-valued diffusion data so that real-valued signals with zero-mean Gaussian noise can be extracted.

### *4.3. Future Directions*

In DMRI, patch-based methods have a wide range of applications, including denoising, atlas building, interpolation, and registration. However, existing patch-based methods in diffusion MRI define patches in the $x$-space. In the future, we will extend our $q$-space patch matching strategy to these applications to cater to the directional nature of diffusion MRI data.

## 5. Conclusion

In this paper, we have proposed an improved NLM algorithm that caters to the spatio-angular characteristics of DMRI data. Our method, called XQ-NLM, performs patch matching in $x$-$q$ space, allowing information from highly curved white matter structures to be used for effective noise removal. Extensive experiments demonstrate that XQ-NLM improves the SNRs of DW images, preserves structural details, and reduces spurious fiber peaks that result from noise.

## Acknowledgment

This work was supported in part by NIH grants (NS093842, EB022880, EB006733, EB009634, AG041721, MH100217, and AA012388). Data were provided in part by the Human Connectome Project, WU-Minn Consortium (Principal Investigators: David Van Essen and Kamil Ugurbil; 1U54MH091657) funded by the 16 NIH Institutes and Centers that support the NIH Blueprint for Neuroscience Research; and by the McDonnell Center for Systems Neuroscience at Washington University.